\documentstyle[11pt,aaspp4,flushrt,epsfig]{article}
\newcommand{\ab}{{_{{\scriptscriptstyle A,B}}}}
\newcommand{\da}{{{\delta}_{{\scriptscriptstyle A}}}}
\newcommand{\done}{{{\delta}_{{\scriptscriptstyle 1}}}}
\newcommand{\dtwo}{{{\delta}_{{\scriptscriptstyle 2}}}}
\newcommand{\ma}{{m_{{\scriptscriptstyle A}}}}
\newcommand{\mone}{{m_{{\scriptscriptstyle 1}}}}
\newcommand{\mtwo}{{m_{{\scriptscriptstyle 2}}}}
\newcommand{\pab}{{p_{{\scriptscriptstyle AB}}}}
\newcommand{\phionetwo}{{\Phi_{{\scriptscriptstyle 12}}}}
\newcommand{\ponetwo}{{p_{{\scriptscriptstyle 12}}}}
\newcommand{\rone}{{{\vec r}_{{\scriptscriptstyle 1}}}}
\newcommand{\ro}{\scriptscriptstyle {\rho}}
\newcommand{\rtwo}{{{\vec r}_{{\scriptscriptstyle 2}}}}
\newcommand{\ronei}{{r_{{\scriptscriptstyle 1i}}}}
\newcommand{\rtwoj}{{r_{{\scriptscriptstyle 2j}}}}
\newcommand{\rAj}{{r_{{\scriptscriptstyle Aj}}}}
\newcommand{\rv}{{\vec r}}
\newcommand{\ra}{{{\rv}_{{\scriptscriptstyle A}}}}
\newcommand{\rb}{{{\rv}_{{\scriptscriptstyle B}}}}
\newcommand{\rh}{{\hat r}}
\newcommand{\rhone}{{{\hat r}_{{\scriptscriptstyle 1}}}}
\newcommand{\rhtwo}{{{\hat r}_{{\scriptscriptstyle 2}}}}

\newcommand{\rha}{{{\hat r}_{{\scriptscriptstyle A}}}}
\newcommand{\rhb}{{{\hat r}_{{\scriptscriptstyle B}}}}
\newcommand{\RoA}{{{\rho}_{{\scriptscriptstyle A}}}}
\newcommand{\Rone}{{{\rho}_{{\scriptscriptstyle 1}}}}
\newcommand{\Rotwo}{{{\rho}_{{\scriptscriptstyle 2}}}}
\newcommand{\sA}{{s_{{\scriptscriptstyle A}}}}
\newcommand{\sB}{{s_{{\scriptscriptstyle B}}}}
\newcommand{\sone}{{s_{{\scriptscriptstyle 1}}}}
\newcommand{\stwo}{{s_{{\scriptscriptstyle 2}}}}
\newcommand{\sthree}{{s_{{\scriptscriptstyle 3}}}}
\newcommand{\sfour}{{s_{{\scriptscriptstyle 4}}}}
\newcommand{\se}{{\sigma_{{\scriptscriptstyle 8}}}}
\newcommand{\va}{{{\vec v}_{{\scriptscriptstyle A}}}}
\newcommand{\vs}{{v_{{\scriptscriptstyle 12}}}}
\newcommand{\vg}{{v_{{\scriptscriptstyle 12g}}}}
\newcommand{\vv}{{{\vec v}_{{\scriptscriptstyle 12}}}}
\newcommand{\vt}{{{\tilde v}_{{\scriptscriptstyle 12}}}}
\newcommand{\vone}{{{\vec v}_{{\scriptscriptstyle 1}}}}
\newcommand{\vtwo}{{{\vec v}_{{\scriptscriptstyle 2}}}}
\newcommand{\xb}{\bar{\xi}}

\newcommand{\xbb}{\bar{\hspace{-0.08cm}\bar{\xi}}}

\slugcomment{CfPA/98-th-17}
\lefthead{Ferreira, Juszkiewicz, Feldman, Davis, \& Jaffe }
\righthead{Large scale flows}

\begin{document}
\tighten
\title{Streaming velocities as a dynamical estimator of $\Omega$}
\author{
P. G. Ferreira\altaffilmark{1},
R. Juszkiewicz\altaffilmark{2,3,4,5},
H. A. Feldman\altaffilmark{2,3},
M. Davis\altaffilmark{1},
and A. H. Jaffe\altaffilmark{1}
}
\altaffiltext{1}{Center for Particle Astrophysics and Astronomy Department,
University of California, Berkeley, CA94720, USA}
\altaffiltext{2}{D{\'e}partement de Physique Th{\'e}orique, Universit{\'e}
de Gen{\`e}ve, CH-1211 Gen{\`e}ve, Switzerland}
\altaffiltext{3}{Department of Physics and Astronomy, University of Kansas,
Lawrence, KS 66045}
\altaffiltext{4}
{On leave from
Copernicus Astronomical Center, 00-716 Warsaw, Poland}
\altaffiltext{5}{Joseph Henry Laboratories,
Princeton University,
Princeton NJ 08544}

\begin{abstract}
It is well known that estimating
the mean pairwise velocity of galaxies,  
$\vs$, from the
redshift space galaxy correlation function is difficult
because this method is highly sensitive
to the assumed model of the pairwise velocity
dispersion. Here we propose an
alternative method to estimate $\vs$ directly from peculiar velocity
samples, which contain redshift-independent distances as well as galaxy
redshifts.
In contrast to other dynamical
measures which determine $\beta\equiv\Omega^{0.6}\se$, this method
can provide an estimate of $\Omega^{0.6}\se^2$
for a range of $\se$
where $\Omega$ is the cosmological density parameter, while $\se$
is the standard normalization for the power spectrum of density
fluctuations. We demonstrate how to measure this quantity from
realistic catalogues.

\end{abstract}


\tighten
\section{Introduction}

In this {\it Letter} we investigate the possibility of using the
``mean tendency of well-separated galaxies to approach each other''
(Peebles 1980, hereafter LSS) to measure the cosmological
density parameter, $\Omega$. The statistic we consider is the mean relative
pairwise velocity of galaxies, $\vs$. It was introduced
in the context of the BBGKY theory (Davis and Peebles 1977),
describing the dynamical evolution of a collection of particles
interacting through gravity. In this discrete picture,
$\vv$ is defined as the mean value of the peculiar
velocity difference of a particle pair at separation ${\vec r}$
(LSS, Eq. 71.4). In the fluid limit, its analogue is the
pair-density weighted relative velocity
(\cite{fisher94,rj98a}),
\begin{eqnarray}
\vv(r) \; = \; \langle \, \vone - \vtwo \, \rangle_{\ro} \; = \;
{{ \langle(\vone -
\vtwo ) (1 + \done)(1 +\dtwo ) \rangle}
\over {1 \; + \; \xi(r)}} \; ,
\label{def}
\end{eqnarray}
where $\, \va \,$ and $\, \da = \RoA / \langle \rho \rangle - 1 \,$
are the peculiar velocity and fractional density contrast
of matter at a point $\ra$, $r=|\rone - \rtwo|$, and $\xi(r) =
\langle\done\dtwo\rangle$ is the two-point correlation function.
The pair-weighted average, $\langle\cdots\rangle_{\ro}$,
differs from simple spatial averaging, $\langle\cdots\rangle$,
by the weighting factor
$\Rone\Rotwo \,\langle \Rone\Rotwo\rangle^{-1}$,
proportional to the number-density of particle pairs.
In gravitational instability theory,
the magnitude of $\vv(r)$ is related to the two point correlation
function, $\xi(r)$, through the pair conservation equation
(LSS, Eq. 71.6).
For models with Gaussian initial conditions, the
solution of the pair conservation equation
is well approximated by (\cite{rj98})
\begin{eqnarray}
\vs (r) \; &=& \; - \, {\textstyle{2\over 3}} \, H r f \,
\xbb (r)[1 + \alpha \; \xbb (r)] \;,
\label{2nd-order}\\
\xb(r) \; &=& \; (3/r^3)\, \int_0^r \, \xi(x) \, x^2 \, dx \; 
\equiv \; \xbb(r) \, [\, 1 + \xi(r) \, ]
\label{xb}
\end{eqnarray}
Here
$\alpha$ is a parameter, which depends on the logarithmic slope
of $\xi(r)$, while
$f = d \, \ln D/d \ln a$, with
$D(a)$ being the standard linear growing mode
solution and $a$ -- the cosmological
expansion factor (see e.g., LSS, \S 11). 
Finally, $H = 100\;h\;\mbox{km}\;\mbox{s}^{-1}\;\mbox{Mpc}$ is 
the present value of the Hubble constant.
For a wide class of cosmological
models, including those with non-zero cosmological constant,
$f \approx \Omega^{0.6}$ (\cite{jim93}, hereafter PPC, \S 13).
If $\xi \propto r^{-\gamma}$, and
$\gamma$ is scale-independent, the predicted streaming velocity can
be expressed in terms of $\Omega$ and the standard normalization
parameter, $\se$ -- the rms matter density contrast in a ball
of radius $8h^{-1}\;\mbox{Mpc}$. For a pure
power-law $\xi(r)$, we have (PPC, Eq. 7.72)
\begin{equation}
\xb(r) \; = \; 3\,\xi(r)\,/\,(3-\gamma) \;
= \; \se^2 (16\;h^{-1}\;\mbox{Mpc}/r)^{\gamma}(4-\gamma)(6-\gamma)/24 \; .
\label{xi-xibar}
\end{equation}
For $0 < \gamma < 3$, the parameter $\alpha$
is given by (\cite{rj98})
\begin{equation}
\alpha \; \approx \; 1.2 \; - \; 0.65 \, \gamma \; .
\label{alpha}
\end{equation}
The approximate solution of the pair conservation equation,
given by equations (\ref{2nd-order}) - (\ref{alpha}) accurately
reproduces results of high resolution N-body simulations
in the entire dynamical range (\cite{rj98}). This approximate
solution was designed to reproduce the second-order Eulerian perturbation
theory solution in the weakly nonlinear regime
($\, |\xi| < 1 \,$) and the stable clustering solution in the strongly
nonlinear regime $(\, \xi \gg 1, \;\; \vs(r) = - Hr\,$; 
see \cite{rs96} and \cite{el96} for the 
second-order correction for $\xi$ and
LSS, \S71 for the stable clustering). 

To get a better idea of how the Equation~(\ref{2nd-order}) can
be used to estimate $\Omega$, let us consider a numerical example:
$\vs$ at a separation $\, r = 10 h^{-1}\, \mbox{Mpc} \,$.
One can use the APM catalogue of galaxies (Efstathiou 1996) for an
estimate of $\gamma$. The slope at the separation considered
is $\gamma=1.75\pm0.1$ (the errors we quote are conservative).
Substituting Eqs.~\ref{xi-xibar} and~\ref{alpha} into
Eq.~\ref{2nd-order}, and setting $\gamma = 1.75$, we get
\begin{equation}
\vs(10 h^{-1}\;\mbox{Mpc}) \; =
\; - \, 605 \, {\se}^2 \Omega^{0.6}
\,(1 + 0.43\se^2)\,/(1+ 0.38{\se}^2)^2 \, {\rm km/s}
\; .
\label{10mpc}
\end{equation}
The above relation shows that at $r = 10 h^{-1}\;\mbox{Mpc}$, $\vs$ is
almost entirely determined by the values of two parameters: $\se$ and
$\Omega$. It is only weakly dependent on $\gamma$. 
This dependence is caused by
the $\alpha\; \xbb$ term in Eq.~\ref{2nd-order}.  However, for all realistic
values of $\gamma$, $\alpha$ is a small number. The uncertainties in the
observed $\gamma$ lead to an error in Eq.~\ref{10mpc} of less than
$10\%$ for $\se\le 1$.

The approximate solution, given by Equation~\ref{2nd-order}
accurately reproduces
$\vs(r)$ curves for dark matter particles,
measured from high resolution N-body simulations
(\cite{rj98}). Moreover, Eq.~\ref{2nd-order} agrees well with
measurements of mean streaming
velocities of ``galaxies'',
$\, \vg(r) \,$, obtained in recent simulations, which
attempt to take into account non-gravitational
processes like star formation and radiative cooling
(\cite{kauf98}). These simulations 
exhibit {\it clustering} bias, but no {\it velocity} bias
($\, \vg = \vs\, $),  
suggesting that galaxies constitute reliable test particles,
driven by the gravitational field of the true mass distribution
even if the galaxies themselves are biased tracers of mass.

\section{The estimator}

Since we observe only the line-of-sight component of the
peculiar velocity, $\sA = \ra\cdot \va/r \equiv 
\rha\cdot \va$, rather than the
full three-dimensional velocity $\va$, it is not possible
to compute $\vs$ directly. Instead, we propose to use the
mean difference between radial velocities of a pair of
galaxies,
$\langle \, \sone - \stwo \, \rangle_{\ro} \; = \;
\vs \, \rh\cdot(\rhone + \rhtwo)/2$,
where $\rv = \rone - \rtwo$.
To estimate $\vs$, we use the simplest least squares techniques,
which minimizes the quantity
$\chi^2(r) \; = \; \sum_{\ab} \, \left[ (\sA - \sB) - \pab
\,\vt(r)/2 \, \right]^2 \; \;$,
where $\pab \equiv \rh \cdot (\rha + \rhb)$ and
the sum is over all pairs at fixed separation $r = |\ra - \rb|$.
The condition $\partial \chi^2 / \,\partial\vt = 0$ implies
\begin{eqnarray}
\vt (r) \; = \; {
{2 \sum \, (\sA - \sB)\, \pab }\over
{\sum \pab^2}} \;\; .
\label{estimator}
\end{eqnarray}
This estimator is appropriate to be
applied to a point process which will sample an underlying continuous
distribution. The sampling is quantified in terms of the selection
function, $\phi(\rv)$. The continuum limit of Eq.~\ref{estimator}
is then
\begin{eqnarray}
\vt (r)=\frac{2 \int d \mone \,
d \mtwo \, \phionetwo \,
(\sone - \stwo)\ponetwo}
{\int d \mone \,
d \mtwo \, \phionetwo \,
\ponetwo^2} \; ,
\label{estcont}
\end{eqnarray}
with $d \ma = \RoA~d^3 \ra$, and
a two-point selection function given by
$\phionetwo \; = \; \delta_D(|{\rone - \rtwo}| - r)
\, \phi(\rone) \, \phi(\rtwo) \;$,
where $\delta_D$ is the Dirac delta function.
For ease of notation we shall denote the denominator
in Eq.~\ref{estcont} by $W(r)$.  If
we take the ensemble average of Eq.~\ref{estcont} we
 find that $\langle \vt(r) \rangle=\vs (r)$.  Note that,
unlike the estimators for the velocity correlation tensor proposed in
\cite{gorski89}, the ensemble average of the estimator is $\vs (r)$
independent of
the selection function. For an isotropic selection function this
estimator is insensitive to systematic effects such as a bulk flow,
large scale shear and small scale random velocities
(as one might expect from virialized objects).

\begin{figure}
\centerline{\psfig{file=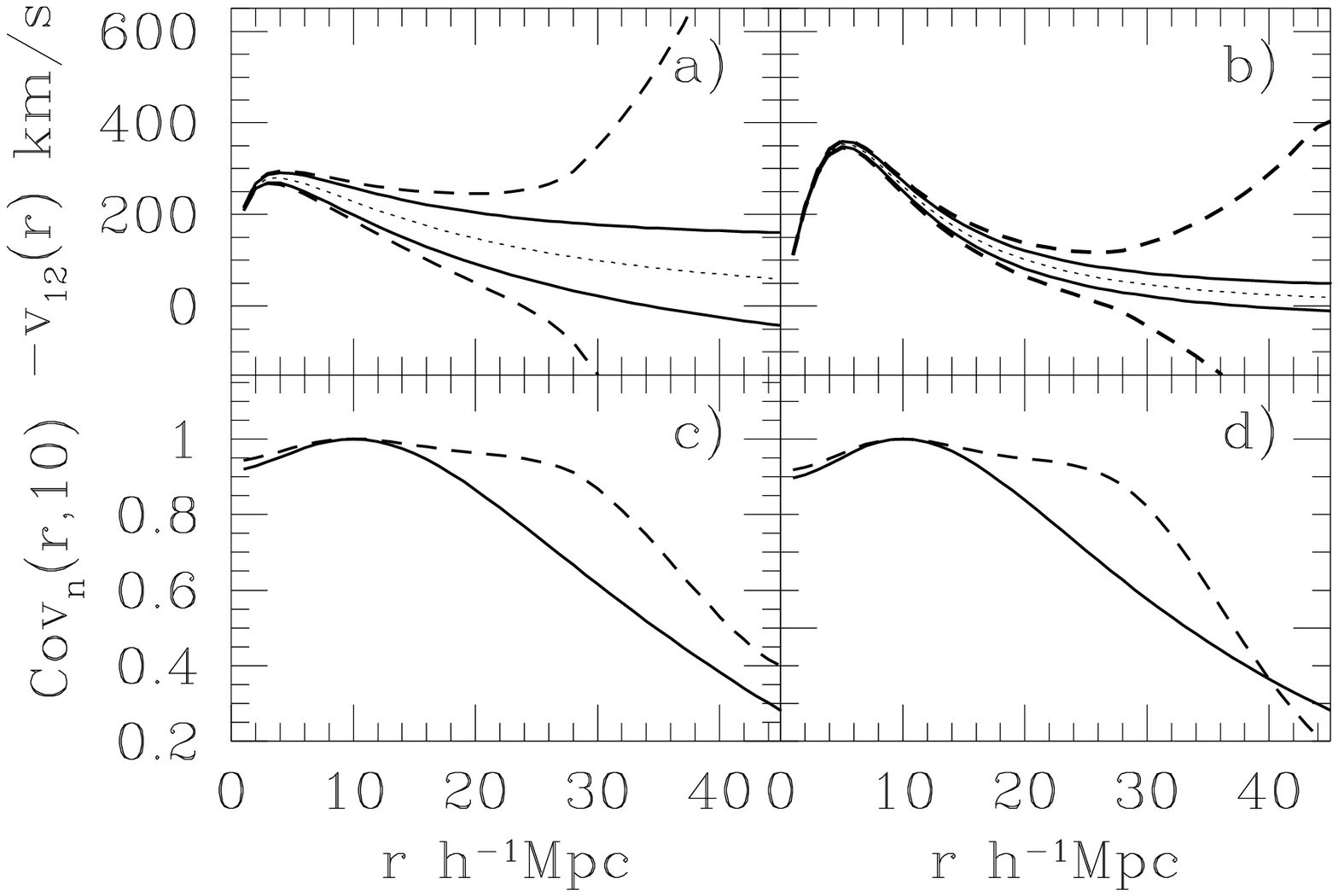,width=12cm}}
\caption{Top panels: $\vs (r)$ (dotted line)  and its variance for both
the full (solid line) and the truncated (dashed line) selection function
for CDM spectra with $\Omega=0.3$ (a) and $\Omega=1$ (b) calculated
using Eq.~\ref{2nd-order} with $\alpha=0$.
The fluctuations are normalized such that $\Omega^{0.6}\se^2=0.49$.
Bottom panels: the normalized covariance function  for CDM models
with $\Omega=0.3$ (c) and $\Omega=1$ (d). The solid (dashed) line corresponds
to the full (truncated) selection function.}
\label{fig1}
\end{figure}

To assess how useful this statistic is in practice we calculate the
covariance matrix of $\vt(r)$; this involves calculating the
ensemble average,
$\langle (\sone-\stwo)(\sthree-\sfour)\rangle_{\ro} =
\langle (\sone-\stwo)(\sthree-\sfour)\rangle + $
higher order terms.
Unlike most statistics, the number weighting leads to
a variance which is of the same order in perturbation theory,
${\cal O}(\delta^2)$, 
as the actual quantity one is
trying to estimate. The covariance between estimates of the pairwise
velocity at two different separations,
$\vt = \vt(r)$, and $\vt' = \vt(r')$, can be expressed as
\begin{eqnarray}
\langle \vt \vt'\rangle-\langle \vt \rangle \langle\vt'\rangle=
\int d^3\rone
d^3\rtwo \phi(\rone)\phi(\rtwo)\mu(r,\rone)\mu(r',\rtwo){\ronei}
\Psi_{ij}(r){\rtwoj} \; ,
\end{eqnarray}
where $\rAj$ is the $j^{\rm th}$
cartesian component of $\ra$ ($A = 1,2; \; j = 1,2,3$),
while $\Psi_{ij}(\rv) \, = \,
\langle v_i(0) v_j(\rv)\rangle$ is the velocity correlation tensor,
and
$\mu(r,\rone) \equiv
2W^{-1}(r)\int d^3\rtwo\phi(\rtwo)\delta_D(|\rtwo-\rone|-r)
\ponetwo$.
In the linear regime, $\Psi_{ij}$ can be expressed in terms of
$P(k)$ --- the power
spectrum of density fluctuations (\cite{gorski88}, \cite{groth89}),
$\Psi_{ij} \, (\rv) \; = \; (H_0^2f^2/8\pi^3) \,
\int \, d^3{\vec k} \, P(k) \,(k_i k_j/k^4)\,\exp(i \vec k \cdot \rv)$.
The form of the selection function will dictate the dependence of the
variance on scale. As one would expect, the smaller the depth of the
selection function, the larger the variance. This is illustrated in
Figure~\ref{fig1}(a,b) where we plot the mean (dotted line) and variance
of $\vt(r)$ for two COBE normalized CDM models and for a
choice of two selection functions. In this {\it Letter} we shall use a
selection function of the form
\begin{eqnarray}
\phi(r)\; \propto \; [r(1+(r/r_{*})^2)]^{-1} \; ; \qquad
r_{*} = 30\;h^{-1}\mbox{Mpc} \; ,
\label{selfu}
\end{eqnarray}
which we shall refer to as the ``full'' selection function (plotted as
the solid line in Figure~\ref{fig1}). In many cases 
galaxy catalogues will have a
sharply decaying selection function beyond a certain scale (Strauss $\&$ 
Willick 95) and it is
therefore useful to check the effect such a feature will have on our
estimator.  We shall do so by considering the selection function of
Eq.~\ref{selfu} truncated at $r = r_{*}$. We refer to the latter as
``truncated'' (dashed line in Figure~\ref{fig1}).

A very important feature of this statistic is the possible 
presence of non-negligible
covariance between values of the estimator at different scales.
The fact that the covariance depends on the velocity correlation
tensor, $\Psi$, will lead to larger covariance than what one might
naively expect: the larger coherence length of this quantity
(as compared to either $\xi(r)$ or $\vs(r)$)
leads to a larger coherence in the covariance matrix and consequently
to larger cross correlations between $\vs$ at different scales.
In Figure~\ref{fig1}(c,d), we plot the appropriately
normalized covariance, ${\rm Cov}_n(\vt,\vt')
=[\langle \vt\vt'\rangle-
\langle\vt\rangle\langle\vt'\rangle]/
\sqrt{\langle\vt^2\rangle\langle
{\vt'}~^2\rangle}$ for a range of separations from
$10h^{-1}\;\mbox{Mpc}$. An open universe has a longer coherence length then
the flat universe, and therefore a stronger covariance; also
we see that for a shallow $\phi(r)$,
the  correlations between the estimates of
$\vt(r)$ will be large.

\section{Tests with mock catalogues}

To test the reliability of the results derived above we now apply our
statistic to mock catalogues extracted from N-body simulations of a
dust-filled universe with $\Omega=1$ and $P(k) \propto 1/k$.  We use one
realization of this model universe in a box which is
$235h^{-1}\;\mbox{Mpc}$ on a side and is normalized to $\se$=0.7. From
this box we extract sets of mock catalogues following the
procedure described in Davis, Nusser
$\&$ Willick (1996) however we emphasize several features.  Small-scale
velocities have been suppressed to $\sim200\;\mbox{km}\;\mbox{s}^{-1}$;
this is not a self-consistent procedure and will lower the amplitude of
Eq.~\ref{2nd-order} by $\sim 10 \%$. In exchange for this relatively
small inaccuracy, our mock catalogues reproduce the observations and
observational errors more faithfully. 
Our mock observers
are centered on particles moving at 600~km~s$^{-1}$ with small local
shear; i.e., resembling conditions in the Local Group. In dense regions,
the redshift fingers of god have been collapsed as is done in the Mark
III (Willick {\it et al} 1997) and IRAS (Fisher {\it et al} 1994) 
catalogues. A typical catalogue will have between 6000 to
11000 galaxies.

\begin{figure}
\centerline{\psfig{file=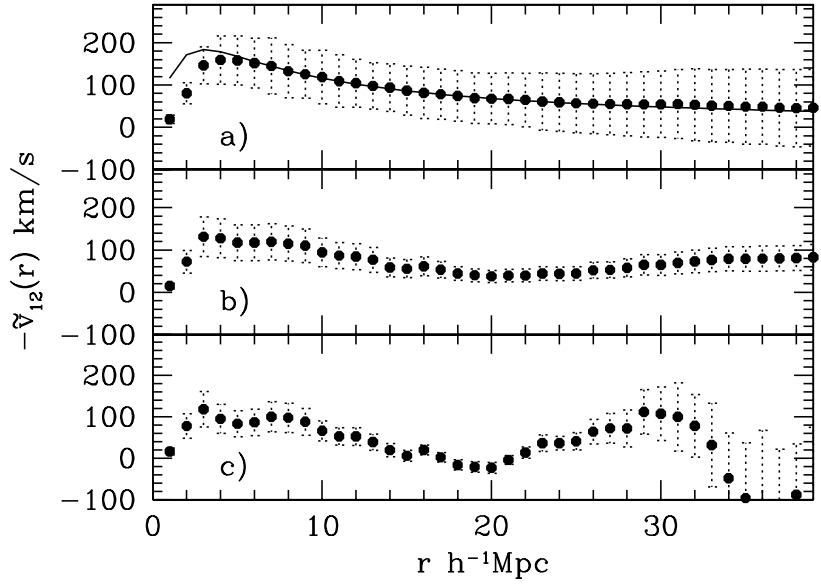,width=12cm}}
\caption{$\vt(r)$ (points) and its variance (dashed lines)
evaluated from mock catalogues described in the text: A) random
observers with the full selection function compared to $\vs (r)$
evaluated from Eq.~\ref{2nd-order} (solid line); b)
A fixed observer with full selection function; c) a fixed observer
with a truncated selection function. The variance is estimated
from the scatter over 20 (a) or 9 (b,c) mock catalogues. }
\label{fig2}
\end{figure}

In Figure~\ref{fig2}(a) we plot $\vt(r)$ with one standard
deviation calculated with 20 mock catalogues extracted with the full
$\phi(r)$ as described in the previous section. Each catalogue has a
different observation position within the simulation volume and so an
average over this set should resemble a true ensemble average. The mean
is consistent with what one would expect from a
direct calculation with Eq.~\ref{2nd-order} (which is plotted in
Figure~\ref{fig2}(a) as a solid line).  We have also performed this
analysis without collapsing the cores; the results changed by very
little. 

We repeat this
calculation for a set of 9 catalogues all constructed from the same
observation point  using the full (Figure~\ref{fig2}b) or truncated 
(Figure~\ref{fig2}c) selection
function to randomly sample a fraction of galaxies within the
simulation box. 
The variance
in $\vt(r)$ is now solely due to finite sampling (``shot noise'');
for catalogues with 2000 to 3000 galaxies we expect the
variance to be $\sqrt{2}$--$\sqrt{3}$ times larger.
We find that truncating the selection function
changes the functional form, or slope, of the mean,
making it a more sharply decreasing
function of $r$ than the ensemble average.  It is therefore crucial when
analyzing a catalogue to restrict oneself to scales much smaller than
the effective cutoff scale of the selection function.

Next we will address the impact of errors in distance measurements
on the estimator, $\vt$. Presently the best estimators
use empirical correlations between intrinsic properties of
the galaxies and luminosities. The errors in such methods
lead to log-normal errors in the estimated distance of around
$20\%$ (for a clear description see Landy $\&$ Szalay 1992).
These errors will naturally lead to biases in cosmological
estimators involving distance measurements and peculiar velocities and
are generically called Malmquist bias. There are formal prescriptions for
correcting for these biases but they rely on assumptions
about the correlations between errors in the distance measurement
and the selection function. Clearly this should be addressed
on a case-by-case basis. For the purpose of this {\it Letter} we shall
assume no correlation between the distance estimator and the
selection function.

We shall model our errors assuming a Tully--Fisher law which
resembles that inferred from the Mark III catalogue.  The line
width, $\eta$, and absolute magnitude, $M$, are related by
$\eta=\epsilon M+\eta_0$ with $\epsilon=-0.1$ and $\eta_0=-2$.
The line width obeys a Gaussian distribution with $\sigma_\eta=0.05$
which lead to log-normal variance in the distance estimator, $d$, of
$\sigma_{\ln d}=23\%$. Using one set of galaxies extracted
from the simulation box we generate
one hundred catalogues with these galaxies  assuming random errors in 
the distance measurement according to the above distribution.
To assess the importance of Malmquist bias we first evaluate
$\vt$ using the raw (uncorrected) distances. To correct for
Malmquist bias we use the prescription put forward in Landy $\&$
Szalay (1992): we correct the raw distance, $d_R$, to get the true
distance, $d_t=d_R\exp(3.5\sigma^2_{\ln d})
\phi_r(\exp(\sigma^2_{\ln d})d_R)/\phi_R(d_R)$, where $\phi_R$
is the selection function estimated from the raw distances.
In principle, given our assumptions, this should correct for
Malmquist bias.

\begin{figure}
\centerline{\psfig{file=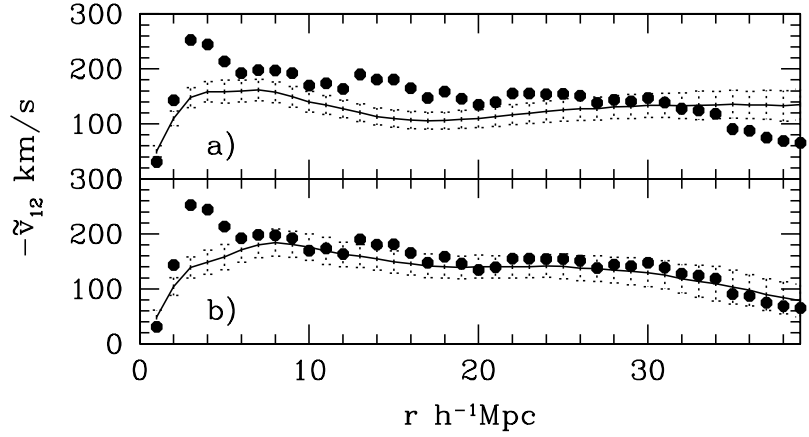,width=12cm}}
\caption{$\vs (r)$  and its variance evaluated from 100 mock catalogues
with errors (described in the text) and the full selection function. The
solid points are the  $\vt$ of the error-free simulation seen from the
same observation point,
the solid line is the mean and dashed lines are the 1 $\sigma$. a)
uncorrected distances; b) distances corrected for Malmquist bias }
\label{fig3}
\end{figure}

\begin{figure}
\centerline{\psfig{file=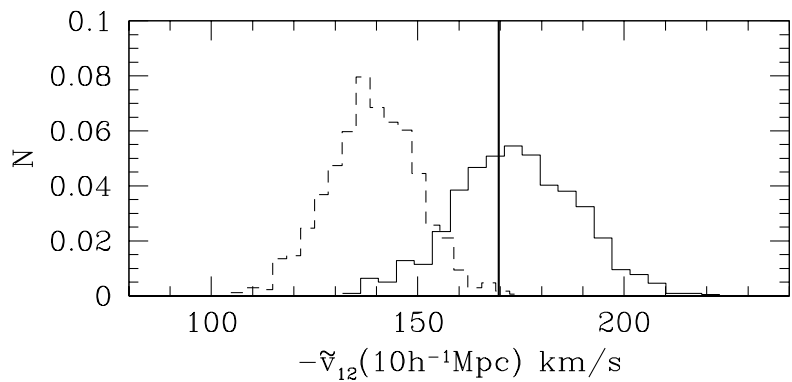,width=12cm}}
\caption{ The effect of generalized Malmquist bias on ${\tilde
v}_{12}(10h^{-1}\;\mbox{Mpc})$. The thick vertical line is the true value,
the dashed  histogram is the uncorrected estimate and the solid
histogram is corrected for Malmquist bias. This is taken from
1000 realizations of log-normal errors (as described in the text).
The histograms have been normalized to unity}
\label{fig4}
\end{figure}

In Figure~\ref{fig3}(a), we plot the results for the uncorrected simulations;
Malmquist errors systematically lower the values of $\vt$
on small scales while enhancing its amplitude on large scales
(where the effect should be more dominant).
However in Figure~\ref{fig3}(b) we show
that with the correction for general Malmquist errors to the distance
estimator, it is possible to overcome this discrepancy. The 1-$\sigma$
errors now encompass the true $\vt$ over a wide range of
scales. The Malmquist effect is more obvious in Figure~\ref{fig4} where we plot
the distribution of $\vt(10h^{-1}\;\mbox{Mpc})$ for 1000
realizations with and without correction for Malmquist errors. If
uncorrected, these errors will induce a bias of up to 30$\%$ in $\vt$
and lead to
an underestimate of $\Omega^{0.6}\se^2$. If properly 
accounted for, one can
see from Figure~\ref{fig4} that this bias can be easily overcome.

\section{Discussion}

In this {\it Letter} we propose to estimate the
mean pairwise streaming velocities of galaxies
directly from peculiar velocity samples. 
We argue that it is a powerful measure of $\Omega^{0.6}\se^2$.
Combined with other dynamical estimates of $\Omega^{0.6}\se$ this
allows a direct estimate of $\Omega$.  Our simulations
show that this method is more
robust than the  $\xi(r_p, \pi)$ analyses
of redshift catalogues (\cite{fisher94} and references therein)
because unlike the redshift space correlation function,
$\vs$ is not sensitive to
the presence of rich clusters of galaxies in the sample. 
Moreover, for $\vs$, the random velocity errors average to zero
instead of adding in quadrature as in the $\xi(r_p, \pi)$ method
which estimates the pairwise velocity dispersion. 

We identified three possible sources of systematic errors in
estimates of $\vs$ made directly from radial peculiar velocities
of galaxies. We also found ways of
reducing these errors; these techniques were successfully tested
with mock catalogues. The potential sources of errors and their
proposed solutions can be summarized as follows.

(1) On the theoretical front, assuming a linear theory
model of $\vs(r)$ at $r \approx
10h^{-1}\;\mbox{Mpc}$ can introduce a considerable systematic error in
the resulting estimate of $\se^2\Omega^{0.6}$.
For example, if $\se = 1$ using the linear prediction for
$\vs$ at $r = 10 h^{-1}\;\mbox{Mpc}$ would introduce a 25\%
systematic error (see eq.~[\ref{10mpc}]).
We solve this problem by using the nonlinear expression
for $\vs$, derived by Juszkiewicz {\it et al.} (1998b).

(2) On the observational front,
a shallow selection function induces a
large covariance between $\vt$ on different scales. This
must be taken into consideration by measuring $\vt(r)$ only
on sufficiently small scales. A rule of thumb is that for
estimating $\vt$ at $10h^{-1}\;\mbox{Mpc}$, the selection function
should be reasonably homogeneous out to at least $30h^{-1}\;\mbox{Mpc}$.

(3) Finally, care must be taken with generalized Malmquist bias due to
log-normal distance errors; these induce a systematic error in $\vt$.
We have shown that, under certain assumptions about selection and
measurement errors, the method of Landy $\&$ Szalay (1992) for corrected
distance estimates allows one to recover the true $\vt$.  Naturally,
this particular correction must be addressed on a case-by-case basis,
given that different data sets will have different selection criteria
and correlations between galaxy position and measurement errors.

In a future publication we shall analyze the Mark III (Willick {\it et
al.} 1997) and the SFI (da Costa {\it et al.} 1996) catalogues of
galaxies with this in mind.

\section*{Acknowledgments}
We thank Jonathan Baker, Stephane
Courteau, Luis da Costa and Jim Peebles,
our referee, for useful comments and suggestions.  
This work was supported in
part by NSF grant AST-95-28340 and NASA grants NAG5-1360 and NAG5-6552
at UCB, by
the NSF-EPSCoR program and the GRF at the University of Kansas,
by the Poland-US M. Sk{\l}odowska-Curie Fund,
by KBN grants No. 2.P03D.008.13 and
2.P03D004.13 in Poland and 
by the Tomalla Foundation in Switzerland.
PGF also thanks JNICT (Portugal).
This work was
conceived in the creative atmosphere of the Aspen Center for Physics,
and we thank the Organizers of the meeting held there in the Summer of
1997.

\eject

\end{document}